\documentclass[aps,prl,twocolumn,preprintnumbers,superscriptaddress,nofootinbib]{revtex4-1}

\usepackage{amsmath}
\usepackage{hyperref}
\usepackage{graphicx}
\usepackage[dvipsnames]{xcolor}

\def\newpar{\vskip4pt}

\def\beq{\begin{equation}}
\def\eeq#1{\label{#1}\end{equation}}
\def\eeqn{\end{equation}}

\def\beqa{\begin{eqnarray}}
\def\eeqa#1{\label{#1}\end{eqnarray}}
\def\eeqan{\end{eqnarray}}
\def\CR{\nonumber \\ }

\def\eqref#1{(\ref{#1})}

\def\L{{\cal L}}
\def\H{{\cal H}}
\def\O{{\cal O}}
\def\M{{\cal M}}

\def\nnbar{$n$-$\bar n$~}
\def\Lnn{\Lambda_{n\bar n}}
\def\Onn{\O_{n\bar n}}
\def\taunn{\tau_{n\bar n}}
\def\LQCD{\Lambda_\text{QCD}}

\begin{document}

\preprint{DESY-18-062}
\preprint{LCTP-18-14}

\title{
Implications of an Improved Neutron-Antineutron Oscillation Search for Baryogenesis: \\ A Minimal Effective Theory Analysis
}



\author{Christophe~Grojean}
\affiliation{Deutsches Elektronen-Synchrotron (DESY), 22607 Hamburg, Germany}
\affiliation{Institut f\"ur Physik, Humboldt-Universit\"at zu Berlin, 12489 Berlin, Germany}
\author{Bibhushan~Shakya}
\affiliation{Department of Physics, University of Cincinnati, Cincinnati, Ohio 45221, USA}
\affiliation{Leinweber Center for Theoretical Physics (LCTP), University of Michigan, Ann Arbor, Michigan 48109, USA}
\author{James~D.~Wells}
\author{Zhengkang~Zhang}
\affiliation{Leinweber Center for Theoretical Physics (LCTP), University of Michigan, Ann Arbor, Michigan 48109, USA}



\begin{abstract}
Future neutron-antineutron ($n$-$\bar n$) oscillation experiments, such as at the European Spallation Source and the Deep Underground Neutrino Experiment, aim to find first evidence of baryon number violation. We investigate implications of an improved $n$-$\bar n$ oscillation search for baryogenesis via interactions of $n$-$\bar n$ mediators, parameterized by an effective field theory (EFT). We find that even in a minimal EFT setup, there is overlap between the parameter space probed by $n$-$\bar n$ oscillation and the region that can realize the observed baryon asymmetry of the universe. The mass scales of exotic new particles are in the TeV-PeV regime, inaccessible at the LHC or its envisioned upgrades. Given the innumerable high energy theories that can match to, or resemble, the minimal EFT that we discuss, future $n$-$\bar n$ oscillation experiments could probe many viable theories of baryogenesis beyond the reach of other experiments.
\end{abstract}


\maketitle



\paragraph{Introduction}---\,
The search for physics beyond the Standard Model (BSM) requires efforts at both high energy and intensity frontiers. In this regard, a particularly powerful probe is offered by rare processes that violate (approximate) symmetries of the Standard Model (SM), such as baryon and lepton numbers ($B$ and $L$), which can be inaccessible to high energy colliders but within reach of low-energy experiments. A well-known example is proton decay, whose non-observation leads to strong constraints on $\Delta B=\Delta L=\pm1$ new physics even at the scale of Grand Unified Theories (GUTs), $\sim 10^{16}$\,GeV~\cite{Miura:2016krn,TheSuper-Kamiokande:2017tit}. 

Baryon and lepton number violation are intricately tied to one of the outstanding puzzles in fundamental physics, the origin of the baryon asymmetry in the universe. If baryogenesis occurs at temperatures above the weak scale, $B-L$ violation is required to avoid washout by electroweak sphalerons. In this regard, constraints from proton decay (which conserves $B-L$) are not applicable. Here we consider instead $B$-violating, $L$-conserving new physics at an intermediate (sub-GUT) scale, so that baryogenesis may proceed both above and below weak scale temperatures. From the low energy point of view, effects of heavy new particles are encoded in higher dimensional operators in an effective field theory (EFT), where $B$-violating, $L$-conserving interactions can appear first at the dimension-nine level~\cite{Kobach:2016ami}, in the form of $|\Delta B|=2$, $\Delta L=0$ operators. In this case, neutron-antineutron ($n$-$\bar n$) oscillation (see~\cite{nnbarReview} for a recent review) is well placed to search for $B$ violating phenomena and shed light on baryogenesis.\footnote{Other $|\Delta B|=2$, $\Delta L=0$ processes include dinucleon decays: $nn\to\pi^0\pi^0$, $pp\to\pi^+\pi^+$, $pn\to\pi^+\pi^0$ probe the same operators as \nnbar oscillation, but with a lower sensitivity at present~\cite{Gustafson:2015qyo}, while $pp\to K^+K^+$~\cite{Litos:2014fxa} can be relevant for $B$-violating new physics with suppressed couplings to first-generation quarks~\cite{Csaki:2011ge,Aitken:2017wie}.}

Current measurements constrain the free neutron oscillation time to be $\taunn \gtrsim 10^8$ s ~\cite{ILL94,SuperK}. Upcoming experiments, in particular at the European Spallation Source (ESS) and also potentially the Deep Underground Neutrino Experiment (DUNE), are poised to improve the reach up to $10^{9\text{-}10}$\,s~\cite{ESS,ESS2,Young:2018talk,Hewes:2017xtr,Barrow:2018talk}. As we will see in detail below, such numbers translate into new physics scales of roughly $(\taunn\LQCD^6)^{1/5}\sim\O(10^{5\text{-}6}\,\text{GeV})$, well above the energies directly accessible at existing or proposed colliders. Discussions of the physics implications of a potential $n$-$\bar n$ oscillation discovery, in particular for baryogenesis, are therefore both important and timely. 

The purpose of this letter is to explore the connection between \nnbar oscillation and baryogenesis in the context of a minimal EFT extension of the SM that realizes direct low scale baryogenesis from $B$ violating decays of new particles mediating \nnbar oscillation. While there exist numerous baryogenesis frameworks, such as electroweak baryogenesis~\cite{Kuzmin:1985mm}, Affleck-Dine baryogenesis \cite{Affleck:1984fy}, and leptogenesis \cite{Fukugita:1986hr} (see e.g.~\cite{Cohen:1993nk,Riotto:1998bt,Riotto:1999yt,Bernreuther:2002uj,Dine:2003ax,Buchmuller:2004nz,Cline:2006ts,Davidson:2008bu,Morrissey:2012db} for reviews), the choice of our minimal EFT is motivated from the bottom-up by imminent improvements in \nnbar oscillation searches. Despite being simplistic, this minimal setup provides a useful template to identify viable baryogenesis scenarios, which may be realized in a similar manner in more complex and realistic theories, that are compatible with an \nnbar oscillation signal within experimental reach (for discussions of some other baryogenesis scenarios that can also involve \nnbar oscillation signals, see e.g.\ \cite{Babu:2006xc, Babu:2006wz, Babu:2008rq, Allahverdi:2010im, Gu:2011ff, Gu:2011fp, Babu:2012vc, Bernal:2012gv, Arnold:2012sd, Babu:2013yca, Allahverdi:2013mza, Patra:2014goa, Herrmann:2014fha, Dev:2015uca, Ghalsasi:2015mxa, Gu:2016ghu, Gu:2017cgp, Calibbi:2017rab, Allahverdi:2017edd}).

\newpar

\paragraph{Improved \nnbar oscillation searches and sensitivity to the scale of new physics}---\,
Neutron-antineutron oscillation has been searched for in the past with both free neutrons~\cite{ILL85,TrigaMark,ILL94} and neutrons bound inside nuclei~\cite{IMB,Kamiokande,Frejus,Soudan,SuperK,SNO}. Among free neutron oscillation searches, the Institut Laue-Langevin (ILL) experiment~\cite{ILL94} sets the best limit to date on the oscillation time, $\taunn > 0.86\times 10^8$\,s at 90\% C.L. Among intranuclear searches, Super-Kamiokande (Super-K)~\cite{SuperK} provides the best limit, which, after correcting for nuclear effects, corresponds to $\taunn > 2.7\times 10^8$\,s at 90\% C.L.\ for the free neutron oscillation time. Improved \nnbar oscillation searches with both free and bound neutrons are under consideration, with sensitivities up to $10^{9\text{-}10}$\,s envisioned at the ESS and DUNE~\cite{ESS,ESS2,Young:2018talk,Hewes:2017xtr,Barrow:2018talk}. 

We now elucidate the connection between $\taunn$ and the new physics scale in the EFT context. The lowest dimension effective operators contributing to \nnbar oscillation at tree level are dimension-nine operators of the form $\Onn\sim(uudddd)$. The classification of these operators dates back to the 1980s~\cite{ChangChang,KuoLove,RaoShrock82,RaoShrock84,nnbarRG} and was refined recently in \cite{BuchoffWagman}, which established an alternative basis more convenient for renormalization group (RG) running. A concise review of the full set of tree-level \nnbar oscillation operators is provided in the Appendix. In what follows, we focus on one of these operators for illustration,
\beqa
\L &\supset& c_1\,\frac{1}{2}\, \epsilon_{ijk}\epsilon_{i'j'k'} (\bar u^c_i P_R d_j) (\bar u^c_{i'} P_R d_{j'}) (\bar d^c_k P_R d_{k'}) +\text{h.c.} \,,\CR
&& \text{with}\;\; c_1\equiv\bigl(\Lnn^{(1)}\bigr)^{-5}\,.
\eeqa{nnbarOperator}
Here $u,d$ are SM up and down quark fields, respectively, and $u^c,d^c$ are their charge conjugates. $i^{(\prime)},j^{(\prime)},k^{(\prime)}$ are color indices, and ``h.c.''~denotes hermitian conjugate. The operator suppression scale $\Lnn^{(1)}$ is generally a weighted (geometric) average of new particle masses, modulo appropriate powers of couplings and loop factors.

If the operator is generated by integrating out new particles at a high scale $M$, computing $\tau_{n\bar n}$ requires RG evolving the EFT down to a low scale $\mu_0$ (usually chosen to be 2\,GeV), where it can be matched onto lattice QCD. The leading contribution to RG rescaling reads~\cite{nnbarRG,BuchoffWagman}
\beqa
\frac{c_1(\mu_0)}{c_1(M)} 
&=& \left[\frac{\alpha_s^{(4)}(m_b)}{\alpha_s^{(4)}(\mu_0)}\right]^{\frac{6}{25}} \left[\frac{\alpha_s^{(5)}(m_t)}{\alpha_s^{(5)}(m_b)}\right]^{\frac{6}{23}} \left[\frac{\alpha_s^{(6)}(M)}{\alpha_s^{(6)}(m_t)}\right]^{\frac{2}{7}} \CR
&=& \bigl\{ 0.726\,,\,\; 0.684\,,\,\; 0.651\,,\,\; 0.624 \bigr\} \,,\CR
&& \text{for}\;\; M = \bigl\{ 10^3\,,\,\; 10^4\,,\,\; 10^5\,,\,\; 10^6 \bigr\} \,\text{GeV} \,.
\eeqa{ZRG}
Here $\alpha_s^{(n_f)}$ is the effective strong coupling with $n_f$ light quark flavors, whose value is obtained with the \texttt{RunDec} package~\cite{RunDec}. Corrections from two-loop running as well as one-loop matching onto lattice QCD operators were recently computed \cite{BuchoffWagman} and are small, and will be neglected in our calculations. No additional operators relevant for \nnbar oscillation are generated from RG evolution.

The $n\to\bar n$ transition rate is determined by the matrix element of the low-energy effective Hamiltonian between the neutron and antineutron states. Thus, once $\langle\bar n|\Onn(\mu_0)|n\rangle$ are known, we can relate $\taunn=\bigl|\langle\bar n|\H_\text{eff}|n\rangle\bigr|^{-1}$ to the six-quark operator coefficients. Recent progress in lattice calculations~\cite{Lattice12,Lattice15} has greatly improved the accuracy and precision on $\langle\bar n|\Onn(\mu_0)|n\rangle$ compared to previous bag model calculations~\cite{RaoShrock82,RaoShrock84} often used in the literature. Using the results in~\cite{Lattice15}, and assuming the operator in Eq.~\eqref{nnbarOperator} gives the dominant contribution to \nnbar oscillation, we can translate the Super-K limit into $\Lnn^{(1)}\gtrsim 4\times 10^5$\,GeV (for a representative RG rescaling factor of 0.7). An improvement on $\taunn$ up to $10^9$ ($10^{10}$, $10^{11}$)\,s will correspond to probing $\Lnn^{(1)}\sim 5\,(8,13)\times 10^5$\,GeV. These numbers are representative of the whole set of \nnbar oscillation operators, and do not vary significantly with the starting point of RG evolution $M$ (see Appendix for details).


\newpar

\paragraph{A minimal EFT for \nnbar oscillation and baryogenesis}---\,
One of the simplest possibilities for generating the operator in Eq.~\eqref{nnbarOperator} at tree level is with a Majorana fermion $X$ of mass $M$ that couples to the SM via a dimension-six operator of the form $\frac{1}{\Lambda^2}Xudd$, which originates at an even higher scale $\Lambda\gg M$ via some UV completion that we remain agnostic about. A familiar scenario that realizes this EFT setup is supersymmetry (SUSY) with $R$-parity violation (RPV), where the bino plays the role of $X$ and the dimension-six operator is obtained by integrating out squarks at a heavier scale. However, this simple EFT with a single BSM state does not allow for sufficient baryogenesis due to unitarity relations: in the absence of $B$-conserving decay channels, $X$ decay cannot generate a baryon asymmetry at leading order in the $B$-violating coupling, a result known as the Nanopoulos-Weinberg theorem~\cite{NWtheorem} (see~\cite{Rompineve} for a recent discussion); meanwhile, $2\to2$ processes $uX\to\bar d\bar d$ and $\bar u X\to dd$ are forced to have the same rate and thus do not violate $CP$.

A minimal extension that can accommodate both $n-\bar n$ oscillation and the observed baryon asymmetry involves two Majorana fermions $X_1, X_2$ (with $M_{X_1}<M_{X_2}$), each having a $B$ violating interaction $\frac{1}{\Lambda^2}Xudd$. In addition, a $B$ conserving coupling between the two is necessary to evade constraints from unitarity relations. In the context of RPV SUSY, this corresponds to the presence of a wino or gluino in addition to the bino, which is known to allow for sufficient baryogenesis~\cite{Cui13,Rompineve,ACN15}.

Guided by minimality, we assume $X_{1,2}$ are both SM singlets, and consider just one of the many possible $B$ conserving operators in addition to the two $B$ violating ones. Our minimal EFT thus consists of the following dimension-six operators that couple $X_{1,2}$ to the SM:\footnote{Our minimal EFT bears similarities with the models studied in~\cite{CheungIshiwata,Baldes1410}. However, these papers focused on baryogenesis using operators of the form $(\bar d^c P_R d)(\bar u^c P_R X)$, which, upon Fierz transformations, are equivalent to generation-antisymmetric components of the $(\bar u^c P_R d)(\bar d^c P_R X)$ operators in Eq.~\eqref{Lag}, and thus do not mediate \nnbar oscillation at tree level.}
\beqa
\L \,&\supset& \,
\eta_{X_1}\, \epsilon^{ijk} (\bar u^c_i P_R d_j) (\bar d^c_k P_R X_1) \CR
&&
+\, \eta_{X_2}\, \epsilon^{ijk} (\bar u^c_i P_R d_j) (\bar d^c_k P_R X_2) \CR
&&
+\, \eta_c\, (\bar u^i P_L X_1) (\bar X_2 P_R u_i) +\text{h.c.}\,, \CR
&&
\text{with}\;\; |\eta_{X_1}|\equiv \Lambda_{X_1}^{-2} \,,\; |\eta_{X_2}|\equiv \Lambda_{X_2}^{-2} \,,\; |\eta_c|\equiv \Lambda_c^{-2} \,.\quad
\eeqa{Lag}
Both $X_1$ and $X_2$ mediate \nnbar oscillation --- 
integrating them out at tree level gives
\beq
c_1 = \frac{1}{\bigl(\Lnn^{(1)}\bigr)^5} = \frac{1}{M_{X_1}\Lambda_{X_1}^4} +\frac{1}{M_{X_2}\Lambda_{X_2}^4}\,.
\eeqn

This setup contains all the necessary ingredients for baryogenesis~\cite{Sakharov}: the Lagrangian in Eq.~\eqref{Lag} violates $B$ and $P$, while nonzero phases of $\eta_{X_1}$, $\eta_{X_2}$, and $\eta_c$ can lead to $CP$ violation; departure from equilibrium can occur in multiple ways, as we discuss below. 
Although a clear simplification, we expect the minimal set of operators in Eq.~\eqref{Lag} to capture the generic qualitative features possible in a two \nnbar mediators setup, which can be realized in more complicated and realistic frameworks.


\newpar
\paragraph{Calculation of the baryon asymmetry}---\,
The relevant processes for baryogenesis include
\begin{itemize}
\item $B$ violating processes: single annihilation $u X_{1,2}\to\bar d \bar d$, $d X_{1,2}\to\bar u \bar d$, decay $X_{1,2}\to udd$, and off-resonance scattering $udd\to\bar u\bar d\bar d$;
\item $B$ conserving processes: scattering $u X_1\to u X_2$, co-annihilation $X_1 X_2\to \bar u u$, and decay $X_2\to X_1 \bar u u$;
\end{itemize}
as well as their inverse and $CP$ conjugate processes. $CP$ violation arises from interference between tree and one-loop diagrams in $u X_{1,2}\leftrightarrow\bar d \bar d$, $u X_1\leftrightarrow u X_2$ and $X_2\leftrightarrow u u d$, and additionally from $udd\leftrightarrow\bar u\bar d\bar d$ (in a way that is related to $X_2\leftrightarrow u u d$ by unitarity). In each case, $CP$ violation is proportional to $\text{Im}(\eta_{X_1}^*\eta_{X_2}\eta_c)\sim\Lambda^{-6}$. We work at leading order in the EFT expansion, i.e.\ $\O(\Lambda^{-4})$ for the rates of $CP$-conserving processes and the $CP$-symmetric components of $CP$-violating processes, and $\O(\Lambda^{-6})$ for the $CP$-violating rates. We choose a mass ratio $M_{X_2}/M_{X_1}=4$, which maximizes $\Gamma(X_2\to udd)-\Gamma(X_2\to \bar u\bar d\bar d)$ for fixed $M_{X_2}$ (see Eq.~\eqref{eGam2}). 

We calculate the baryon asymmetry by numerically solving a set of coupled Boltzmann equations to track the abundances of $X_{1,2}$ and $B-L$ ($B$) above (below) $T=140$\,GeV (we assume sphalerons are active when $T>140$\,GeV, resulting in $Y_B=\frac{28}{79}\,Y_{B-L}$). Our aim is to find regions of parameter space that can achieve the observed $Y_B = 8.6\times10^{-11}$~\cite{Planck13,Planck15}, with suitable choice of $CP$ phases. Technical details of this calculation can be found in the Appendix.

If all three operator coefficients have similar sizes, $\Lambda_{X_1}\sim\Lambda_{X_2}\sim\Lambda_c$, it is difficult to obtain the observed baryon asymmetry in the region of parameter space probed by \nnbar oscillation. For $M_{X_{1,2}}\gtrsim 10^4$\,GeV, the $\Lambda$'s that can be probed are sufficiently low for $X_{1,2}$ to remain close to equilibrium until their abundances become negligible, while efficient washout suppresses $B$($-L$) generation. For lower masses and higher $\Lambda$'s, on the other hand, $X_2$ may freeze out with a significant abundance, and decay out of equilibrium at later times when washout has become inefficient, so that both limitations from the higher mass regime are overcome. However, its $CP$ violating branching fraction $\epsilon_{CP}\sim M_{X_2}^2/\Lambda^2$ is too small to generate the desired $Y_B$. We find that for $\Lambda_{X_1}=\Lambda_{X_2}=\Lambda_c$, the maximum $Y_B$ possible in the ESS/DUNE sensitivity region is $\O(10^{-13})$, well below the observed value.

\begin{figure}[t]
\centering
\includegraphics[width=3.1in]{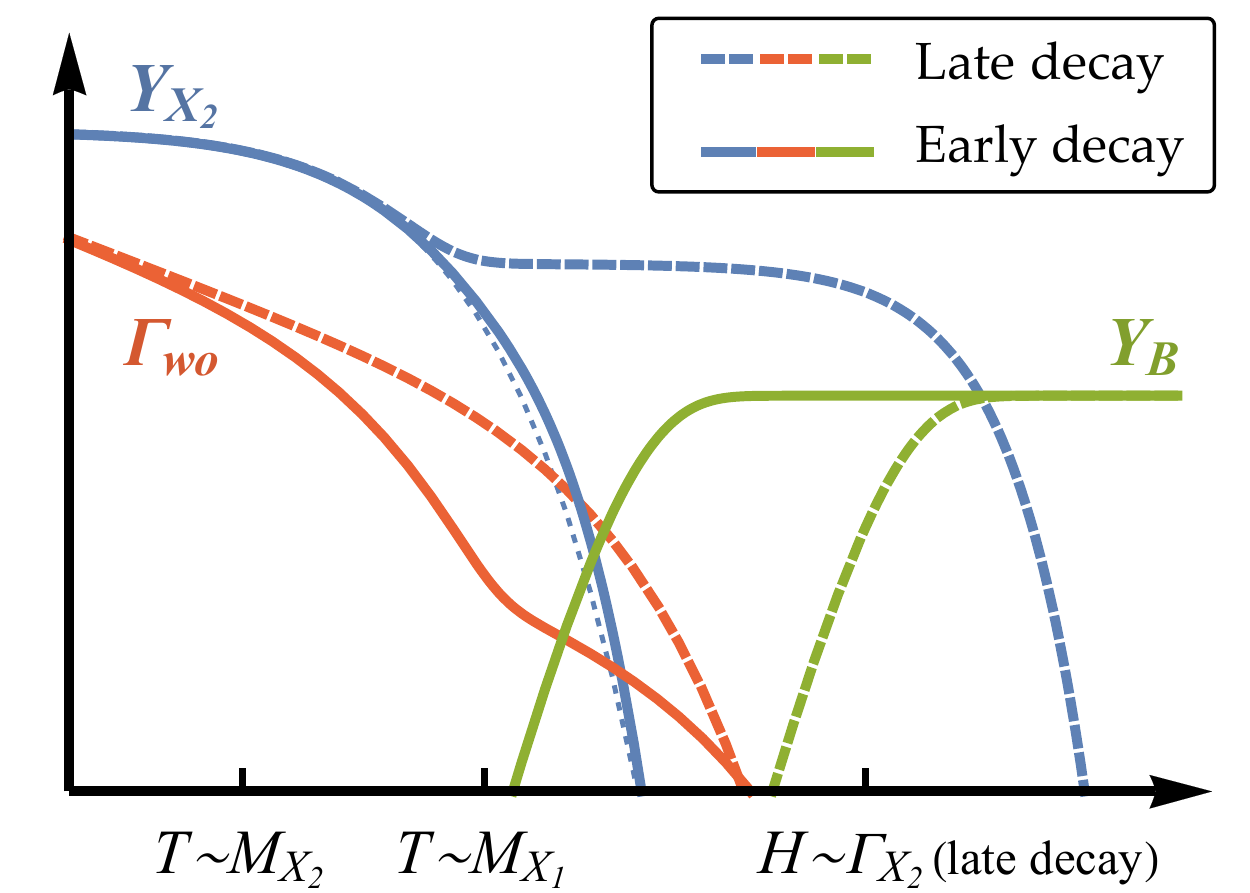}
\caption{Sketches of the evolution of the heavier \nnbar mediator abundance $Y_{X_2}$, washout rate $\Gamma_\text{wo}$ and baryon asymmetry $Y_B$ in the two scenarios considered in this letter (arbitrary normalization). In the late decay scenario, the \nnbar mediator is long-lived and decays out of equilibrium to generate a baryon asymmetry. In the early decay scenario, departure from equilibrium (thin dotted curve) is small, but suppressed washout enables efficient baryogenesis. See text for details.
\label{fig:Cartoon}}
\end{figure}

Achieving the desired baryon asymmetry in the ESS/DUNE reach region therefore requires hierarchical $\Lambda$'s; such scenarios can arise if new particles in the UV theory that mediate the corresponding operators have hierarchical masses and/or couplings, or if the EFT operators are generated at different loop orders. We find compatible regions of parameter space in two distinct scenarios, one with late decays of $X_2$ and the other with earlier decays. These are schematically illustrated in Fig.~\ref{fig:Cartoon}, and discussed in turn below (a detailed analysis with benchmark numerical solutions is presented in the Appendix).

\begin{figure}[t]
\centering
\includegraphics[width=3.1in]{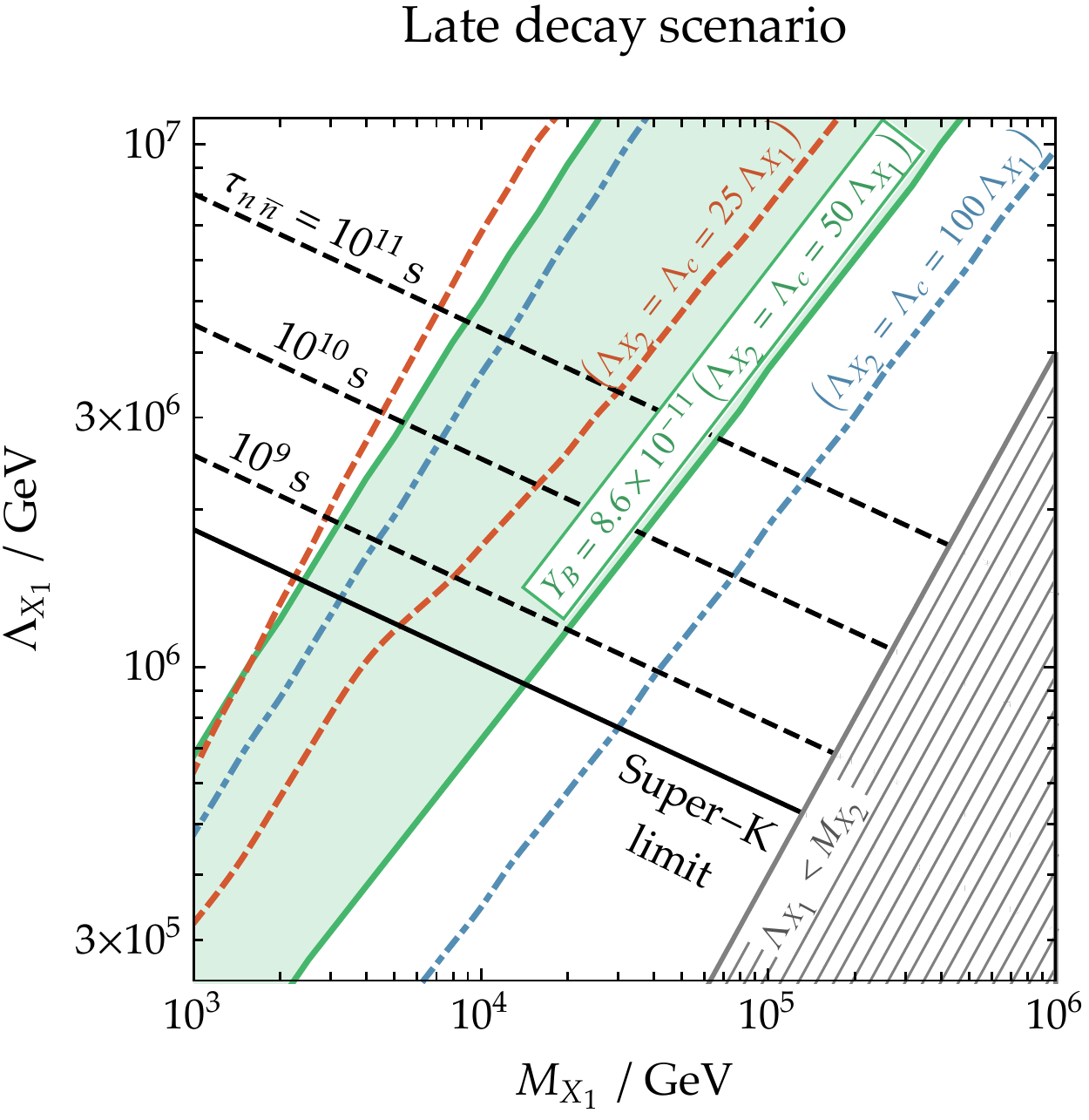}
\caption{
Parameter space of the minimal EFT probed by \nnbar oscillation for the late decay scenario, assuming $M_{X_2}=4\,M_{X_1}$. For $\Lambda_{X_2}=\Lambda_c=50\,\Lambda_{X_1}$, the green shaded region can accommodate $Y_B=8.6\times10^{-11}$. For $\Lambda_{X_2}=\Lambda_c=25\,\Lambda_{X_1}$ ($100\,\Lambda_{X_1}$), viable region is between dashed red (dot-dashed blue) lines. The gray shaded region marks $\Lambda_{X_1}<M_{X_2}$, where EFT validity requires greater than $\O(1)$ coupling.
\label{fig:LateDecay}}
\end{figure}

\begin{figure}[t]
\centering
\includegraphics[width=3.1in]{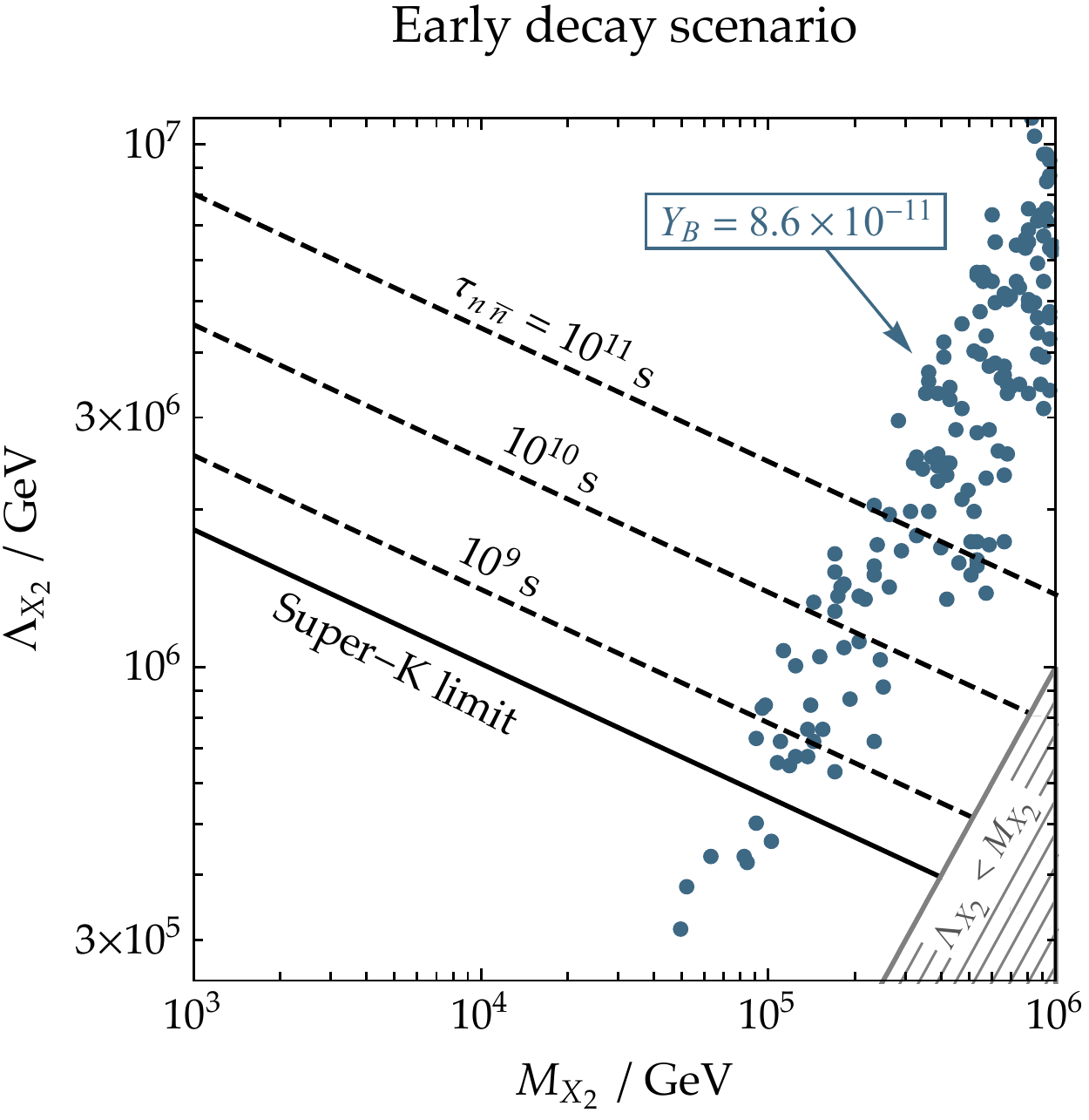}
\caption{
Parameter space of the minimal EFT probed by \nnbar oscillation for the early decay scenario, assuming $M_{X_2}=4\,M_{X_1}$. Points represent solutions with $Y_B=8.6\times10^{-11}$ found in a scan over $\Lambda_{X_2}<\Lambda_{X_1}<100\,\Lambda_{X_2}$, $M_{X_2}<\Lambda_c<\Lambda_{X_2}$. For all these points, $\Lambda_{X_1}\sim10\,\Lambda_{X_2}$ is needed to suppress washout. The gray shaded region marks $\Lambda_{X_2}<M_{X_2}$, where EFT validity requires greater than $\O(1)$ coupling.
\label{fig:EarlyDecay}}
\end{figure}


\newpar

\paragraph{Late decay scenario}---\,
For $\Lambda_{X_2}\sim\Lambda_c\gg\Lambda_{X_1}$, $n$-$\bar n$ oscillation is dominated by $X_1$ exchange and probes the $M_{X_1}$-$\Lambda_{X_1}$ parameter space (see Fig.~\ref{fig:LateDecay}). This hierarchy leads to weaker interactions for $X_2$ compared to the degenerate case, causing it to freeze out with a higher abundance $Y_{X_2}^\text{fo}$. Also, $X_2$ becomes long-lived and decays after washout processes have become ineffective, thereby creating substantial baryon asymmetry (see Fig.~\ref{fig:Cartoon}). In this case, its $CP$-violating branching fraction scales as $\epsilon_{CP}\sim M_{X_2}^2\eta_{X_1}\eta_{X_2} \eta_c/\max(\eta_{X_2}^2,\,\eta_c^2)\sim M_{X_2}^2/\Lambda_{X_1}^{2}$ and does not decouple as $\Lambda_{X_2}$ and $\Lambda_c$ are both increased, enabling $Y_B\sim Y_{X_2}^\text{fo}\,\epsilon_{CP}$ to reach the observed value.

Numerically, we find that this baryogenesis scenario is viable with $\Lambda_{X_2},\,\Lambda_c\gtrsim 20\,\Lambda_{X_1}$ in the parameter space probed by \nnbar oscillation. In Fig.~\ref{fig:LateDecay}, we show regions in the $M_{X_1}$-$\Lambda_{X_1}$ plane that can accommodate the observed baryon asymmetry for various choices of $\Lambda_{X_2}/\Lambda_{X_1}=\Lambda_c/\Lambda_{X_1}$. In each case, the lower boundary of the viable region is effectively determined by the requirement that $X_2$ freezes out with sufficient abundance. As we move upward from this lower boundary, increasing all three $\Lambda$'s while keeping their ratios fixed, at some point we enter a regime where $X_2$ decouples from the SM bath while relativistic, and $Y_{X_2}^\text{fo}$ saturates at $Y_{X_2}^\text{eq}(T\gg M_{X_2})=\frac{1}{\pi^2}\frac{T^3}{s}$, so that further increasing the $\Lambda$'s only reduces $\epsilon_{CP}$ and hence the final $Y_B$. Furthermore, for sufficiently high $\Lambda_{X_2}$ and $\Lambda_c$, $X_2$ dominates the energy density of the universe before it decays (this does not happen for $X_1$ in the parameter space we consider), so that its decay injects significant entropy into the plasma, diluting the baryon asymmetry. Both of these effects -- saturation and dilution -- determine the upper boundary of the viable region.


\newpar

\paragraph{Early decay scenario}---\,
For the opposite hierarchy $\Lambda_{X_1}\gg\Lambda_{X_2}$, \nnbar oscillation is dominated by $X_2$ exchange and probes the $M_{X_2}$-$\Lambda_{X_2}$ parameter space (see Fig.~\ref{fig:EarlyDecay}). In this case, $X_2$ is short-lived, and its abundance closely follows the equilibrium curve. However, small departures from equilibrium, always present in an expanding universe because interaction rates are finite, can be sufficient for baryogenesis if washout can be suppressed. The rates for washout processes involving $X_1$ and $X_2$ are proportional to $n_1\Lambda_{X_1}^{-4}$ and $n_2\Lambda_{X_2}^{-4}$, respectively, where $n_{1,2}$ are the number densities of $X_{1,2}$. If $\Lambda_{X_1}\sim\Lambda_{X_2}$, washout would be efficient until $T\sim M_{X_1}$, i.e.\ until $n_1$ starts to fall exponentially. In contrast, by increasing $\Lambda_{X_1}$, we enter a regime where washout is dominated by $X_2$ processes at high temperatures and becomes inefficient as soon as the temperature falls below $M_{X_2}$ (washout due to $udd\leftrightarrow\bar u\bar d\bar d$, whose rate $\sim T^{11}/M^2\Lambda^8$ falls steeply with $T$, is also irrelevant at this point), resulting in a short period of baryon asymmetry generation from $X_2$ decays (see Fig.~\ref{fig:Cartoon}). Note that increasing $\Lambda_{X_1}$ with respect to $\Lambda_{X_2}$ also helps to increase departures from equilibrium compared to the degenerate case. 

Fig.~\ref{fig:EarlyDecay} shows points in the $M_{X_2}$-$\Lambda_{X_2}$ plane that can realize the observed $Y_B$ through this early decay process, based on a numerical scan over the region $\Lambda_{X_2}<\Lambda_{X_1}<100\,\Lambda_{X_2}$, $M_{X_2}<\Lambda_c<\Lambda_{X_2}$. For the majority of these points, $\Lambda_{X_1}$ is within a factor of two from $10\,\Lambda_{X_2}$, while $\Lambda_c\lesssim 3\,M_{X_2}$. The results can be understood from the competing effects of baryon asymmetry generation and washout, $\Gamma_{\Delta B\neq 0}/\Gamma_\text{wo}\sim M^2\,n_2(\Lambda_{X_1}^2\Lambda_{X_2}^2\Lambda_c^2)^{-1}/(n_1\Lambda_{X_1}^{-4}+n_2\Lambda_{X_2}^{-4})\sim (M^2/\Lambda_c^2)\cdot\min\bigl\{\Lambda_{X_2}^2/\Lambda_{X_1}^2,\,\Lambda_{X_1}^2/\Lambda_{X_2}^2 \,e^{-(M_{X_2}-M_{X_1})/T}\bigr\}$, where the rate of baryon asymmetry generation $\Gamma_{\Delta B\neq 0}$ is calculated from $CP$-violating $X_2$ decays. First of all, a lower ratio $\Lambda_c/M_{X_2}$ is always preferable (within the range of EFT validity), while the ratio $\Lambda_{X_2}/\Lambda_{X_1}$ has an optimal value of $\sim1/10$ as a result of balancing between faster baryon asymmetry generation at higher temperatures (which favors higher $\Lambda_{X_2}/\Lambda_{X_1}$) and later transition to $X_1$-dominated washout (which favors lower $\Lambda_{X_2}/\Lambda_{X_1}$). The requirement of sufficient departure from equilibrium precludes arbitrarily low $\Lambda_c$ and leads to a minimum $M_{X_2}$ for this scenario to work, which we see from Fig.~\ref{fig:EarlyDecay} is a few $\times10^4$\,GeV. Finally, the overall size of $\Lambda_{X_{1,2}}$ is essentially determined by the requirement that $Y_B$ freezes out around the time $\Gamma_{X_2}^{\Delta B\neq 0}/\Gamma_\text{wo}$ reaches its maximum, and is higher for higher $M_{X_2}$.


\newpar

\paragraph{Complementary probes}--- \,
In the region of parameter space that is allowed by existing \nnbar oscillation searches, within reach of the ESS/DUNE, and realizes the observed baryon asymmetry, we find $M_{X_{1,2}}\gtrsim 10^3 (10^4)$ GeV in the late (early) decay scenario. Given that $X_{1,2}$ are SM singlets that only couple to the SM via higher dimensional operators, it is unlikely that they can be detected at the LHC or its envisioned upgrades. Likewise, there are no strong flavor physics constraints on our minimal EFT with just the operators in Eq.\,(\ref{Lag}). We note, however, that this outlook can change in a more complicated model that preserves the general features of baryogenesis of our minimal EFT if at least one of $X_{1,2}$ carries SM charges or couples to other fermion species. For example, colored particles at the TeV scale, such as the gluino in RPV SUSY, could be within LHC reach. Likewise, extending the exotic fermion couplings to other quark flavors can introduce potential constraints from flavor violation considerations such as $K^0-\bar{K}^0$ mixing \cite{CheungIshiwata}. Nevertheless, our minimal EFT study illustrates that \nnbar oscillation might be uniquely placed to probe  realistic baryogenesis scenarios that are inaccessible via other searches.


\paragraph{Conclusions}---\,
Establishing baryon number violation (or the absence thereof up to a certain scale) will have far-reaching implications on our understanding of fundamental particle interactions, in particular on the mechanism that generates the observed baryon asymmetry in our universe. Motivated by the unprecedented sensitivity to \nnbar oscillation that can be achieved at future facilities, the ESS and DUNE in particular, which offers new opportunities to probe $|\Delta B|=2$, $\Delta L=0$ interactions, we studied implications of a potential discovery for baryogenesis scenarios involving \nnbar mediators. We took a bottom-up EFT approach with a minimal set of four-fermion operators coupling the \nnbar mediators to the SM, which, despite being simplistic, sets a useful template that more sophisticated theories can build upon. We identified two viable baryogenesis scenarios -- one involving late out-of-equilibrium decays of a heavy Majorana fermion, and another involving earlier decays assisted by a suppressed washout rate --  that can be realized in the parameter space to be probed by future \nnbar oscillation searches, with no corresponding collider or flavor signals. These results highlight the capability of \nnbar oscillation experiments to probe an important BSM phenomenon, that of baryogenesis, beyond the scope of other searches.


\newpar
\newpar
\begin{acknowledgments}
We thank J.~Barrow, G.~Brooijmans and C.~Cs\'aki for helpful comments and discussions. C.G.\ was supported by the European Commission through the Marie Curie Career Integration Grant 631962, and by the Helmholtz Association. B.S.\ was partially supported by the NSF CAREER grant PHY-1654502 and thanks the CERN and DESY theory groups, where part of this work was conducted, for hospitality. The work of J.D.W.\ was supported by the DoE grant DE-SC0007859 and the Humboldt Research Award. The work of Z.Z.\ was supported by the DoE grant DE-SC0007859, the Rackham Dissertation Fellowship, and the Summer Leinweber Research Award. J.D.W.\ and Z.Z.\ also thank the DESY theory group for hospitality. This work was performed in part at the Aspen Center for Physics, which was supported by National Science Foundation grant PHY-1066293. 
\end{acknowledgments}







\onecolumngrid
%

\vspace{40pt}
\section{Appendix: supplemental material}
\renewcommand{\theequation}{A.\arabic{equation}}
\setcounter{equation}{0}

\subsection*{1.~Neutron-antineutron oscillation operators}

Here we briefly review the effective operator analysis of \nnbar oscillation. Since multiple operators may be present in addition to the representative operator we considered in the letter, to gain intuition about the new physics scale being probed, let us define
\beq
\taunn^{-1} = \bigl|\langle\bar n|\H_\text{eff}|n\rangle\bigr| \equiv \frac{\LQCD^6}{\Lnn^5} \,.
\eeq{Lnndef}
As we will see explicitly below, $\Lnn$ defined here roughly coincides with suppression scales of dimension-nine operators mediating \nnbar oscillation. This is because the nuclear matrix elements $\langle\bar n|\Onn|n\rangle\sim\O(\LQCD^6)$. Taking $\LQCD=180$\,MeV, we have
\beqa
\Lnn &=& 4.25\times10^5\,\text{GeV}\,\biggl(\frac{\taunn}{2.7\times10^8\,\text{s}}\biggr)^{1/5} \label{Lnnreach-current}\\
&=& 5.53\times10^5\,\text{GeV}\,\biggl(\frac{\taunn}{10^9\,\text{s}}\biggr)^{1/5} 
=\, 8.76\times10^5\,\text{GeV}\,\biggl(\frac{\taunn}{10^{10}\,\text{s}}\biggr)^{1/5}
=\, 1.39\times10^6\,\text{GeV}\,\biggl(\frac{\taunn}{10^{11}\,\text{s}}\biggr)^{1/5} \,,\label{Lnnreach-future}
\eeqan
where the number in Eq.~\eqref{Lnnreach-current} shows the current best limit from Super-K.

There are 12 independent operators that contribute to \nnbar oscillation at tree level. Using the basis of~\cite{BuchoffWagman}, we write
\beq
\L_\text{eff} \supset \sum_{i=1}^{6} c_i\O_i +\bar c_i\bar\O_i +\text{h.c.} \,,
\eeqn
where
\beqa
\O_1 &=& \frac{1}{2}\,\epsilon_{ijk}\epsilon_{i'j'k'} (\bar u^c_i P_R d_j) (\bar u^c_{i'} P_R d_{j'}) (\bar d^c_k P_R d_{k'}) \,,\CR
\O_2 &=& \epsilon_{ijk}\epsilon_{i'j'k'} (\bar u^c_i P_L d_j) (\bar u^c_{i'} P_R d_{j'}) (\bar d^c_k P_R d_{k'}) \,,\CR
\O_3 &=& \frac{1}{2}\,\epsilon_{ijk}\epsilon_{i'j'k'} (\bar u^c_i P_L d_j) (\bar u^c_{i'} P_L d_{j'}) (\bar d^c_k P_R d_{k'}) \,,\CR
\O_4 &=& \epsilon_{ijk}\epsilon_{i'j'k'} (\bar u^c_i P_R u_{i'}) (\bar d^c_j P_L d_{j'}) (\bar d^c_k P_L d_{k'}) \,,\CR
\O_5 &=& \bigl(\epsilon_{ijk}\epsilon_{i'j'k'} +\epsilon_{i'jk}\epsilon_{ij'k'}\bigr) (\bar u^c_i P_R d_{i'}) (\bar u^c_j P_L d_{j'}) (\bar d^c_k P_L d_{k'}) \,,\CR
\O_6 &=& \epsilon_{ijk}\epsilon_{i'j'k'} (\bar u^c_i P_L u_{i'}) (\bar d^c_j P_L d_{j'}) (\bar d^c_k P_R d_{k'}) \CR
&& + \bigl(\epsilon_{ijk}\epsilon_{i'j'k'} +\epsilon_{i'jk}\epsilon_{ij'k'}\bigr) (\bar u^c_i P_L d_{i'}) (\bar u^c_j P_L d_{j'}) (\bar d^c_k P_R d_{k'}) \,,
\eeqa{Olist}
and $\bar\O_i$ is obtained by exchanging $P_L\leftrightarrow P_R$ in $\O_i$. Note that since QCD conserves parity, $\O_i$ and $\bar\O_i$ have identical nuclear matrix elements and anomalous dimensions (neglecting weak interactions). Our labeling of $\O_{1,2,3}$ is in accordance with~\cite{BuchoffWagman}, while our $\O_{4,5,6}$ are proportional to their $Q_{5,6,7}$, respectively (their $Q_4$, which we have skipped here, has zero nuclear matrix element).

The operator basis of Eq.~\eqref{Olist} is particularly convenient because different operators do not mix as they are evolved from some high scale(s) $\mu_{(i)}$ down to $\mu_0=2$\,GeV, where lattice calculations of nuclear matrix elements are reported. We have
\beqa
\Lnn^{-5} &=& \biggl|\sum_{i=1}^6 \frac{\langle\bar n|\O_i(\mu_0)|n\rangle}{\LQCD^6} \bigl[c_i(\mu_0) +\bar c_i(\mu_0)\bigr] \biggr| \CR
&=& \Biggl|\sum_{i=1}^6 \frac{\langle\bar n|\O_i(\mu_0)|n\rangle}{\LQCD^6} \Biggl\{\Biggl[\frac{\alpha_s^{(4)}(m_b)}{\alpha_s^{(4)}(\mu_0)}\Biggr]^{\frac{3}{50}} \Biggl[\frac{\alpha_s^{(5)}(m_t)}{\alpha_s^{(5)}(m_b)}\Biggr]^{\frac{3}{46}} \Biggl[\frac{\alpha_s^{(6)}(\mu)}{\alpha_s^{(6)}(m_t)}\Biggr]^{\frac{1}{14}}\Biggr\}^{\gamma_i^{(0)}} 
\bigl[ c_i(\mu) +\bar c_i(\mu)\bigr] \Biggr| \,,
\eeqan
where $\gamma_i^{(0)}$ is the leading order anomalous dimension of operator $\O_i$~\cite{nnbarRG,BuchoffWagman}. Using the latest lattice results in~\cite{Lattice15}, we obtain numerically
\beqa
\Lnn^{-5} 
&=& \Bigl| 0.760(94) \bigl(r_1^{2/7} c_1+\bar r_1^{2/7} \bar c_1\bigr) -4.77(55) \bigl(r_2^{-2/7} c_2+\bar r_2^{-2/7} \bar c_2\bigr) +1.08(10) \bigl(c_3+\bar c_3\bigr) \CR
&& 
-0.0498(61) \bigl(r_4^{6/7} c_4+\bar r_4^{6/7} \bar c_4\bigr) +0.0249(30) \bigl(r_5^{6/7} c_5+\bar r_5^{6/7} \bar c_5\bigr) -0.0249(31) \bigl(r_6^{6/7} c_6+\bar r_6^{6/7} \bar c_6\bigr) \Bigr|\,.
\eeqa{Lnnnum}
Here we have chosen $\mu=10^5$\,GeV as a reference scale to compute the numbers, and introduced $r_i\equiv\alpha_s(\mu_i)/\alpha_s(10^5\,\text{GeV})$, $\bar r_i\equiv\alpha_s(\bar\mu_i)/\alpha_s(10^5\,\text{GeV})$ to account for effects due to different choices (when $\O_i$ and $\bar\O_i$ are renormalized at $\mu_i$ and $\bar\mu_i$, respectively, rather than at $10^5$\,GeV).

\begin{figure}[t]
\centering
\includegraphics[width=3.5in]{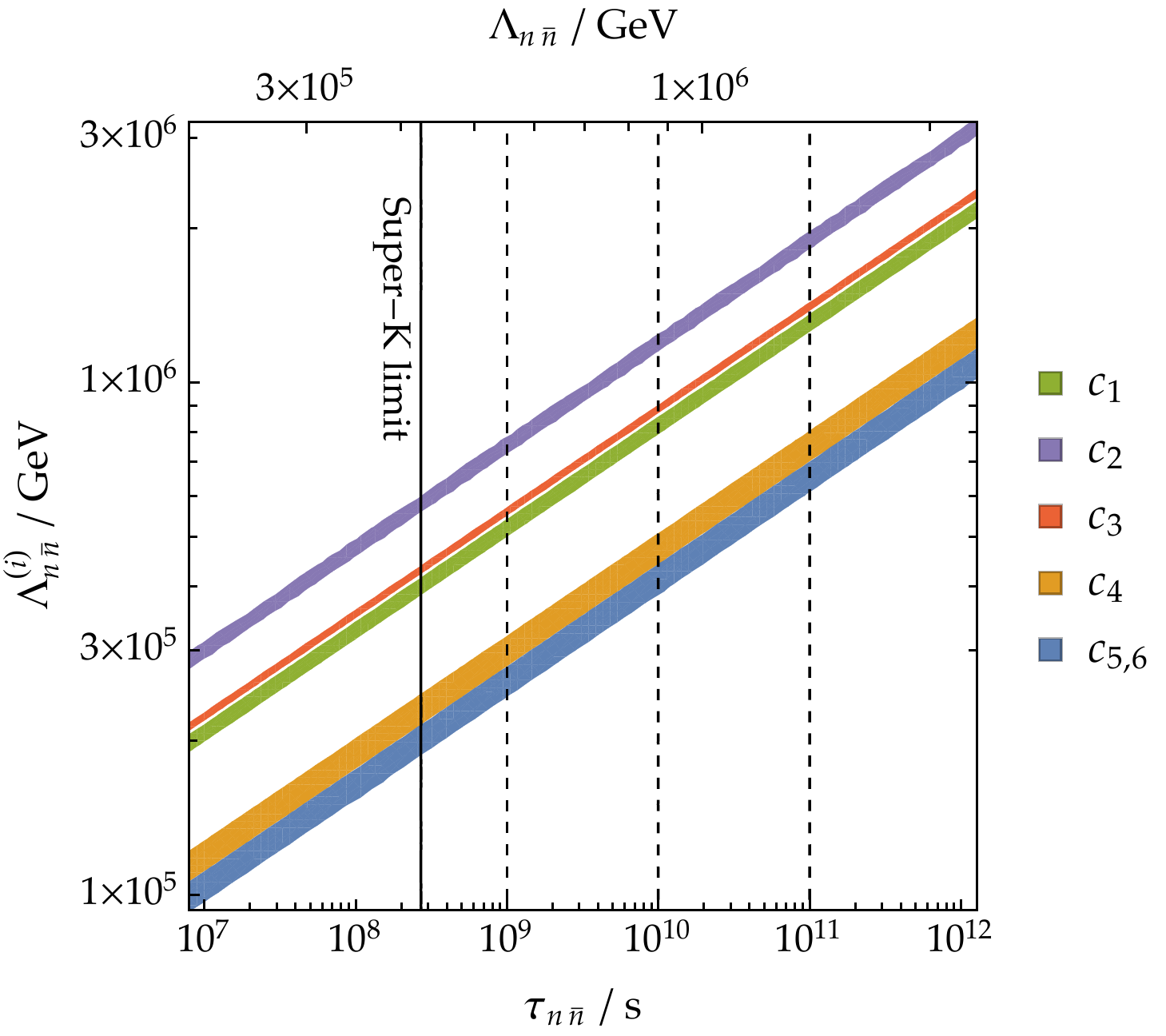}
\caption{\label{fig:Lnn}
Suppression scale $\Lnn^{(i)}\equiv c_i^{-1/5}$ of the $|\Delta B|=2$ six-quark operators $\O_i$ in Eq.~\eqref{Olist} that can be probed with free neutron oscillation time $\taunn$ (corresponding to new physics scale $\Lnn\equiv(\taunn\LQCD^6)^{1/5}$ as defined in Eq.~\eqref{Lnndef}) when each operator is considered individually. The widths of the bands arise from variations of $\langle\bar n|\O_i(\mu_0)|n\rangle$ within current lattice calculation uncertainties, and of the starting point of RG evolution $\mu_i$ between $10^3$\,GeV and $10^6$\,GeV. The results apply equally to the parity-conjugate operators $\bar\O_i$. Existing and future \nnbar oscillation searches are sensitive to $\Lnn^{(i)}\sim\O(10^{5\text{-}6}\,\text{GeV})$.
}
\end{figure}

In the special case that the RHS of Eq.~\eqref{Lnnnum} is dominated by a single term, say the one proportional to $c_i\equiv\bigl(\Lnn^{(i)}\bigr)^{-5}$, we can establish a correspondence between $\taunn$ (equivalently $\Lnn$) and $\Lnn^{(i)}$. This is shown in Fig.~\ref{fig:Lnn}. As mentioned above, all $\Lnn^{(i)}$'s are close to the universal $\Lnn$ defined in Eq.~\eqref{Lnndef}. Among them, $\Lnn^{(4,5,6)}$ are somewhat lower because the corresponding operators have larger (positive) anomalous dimensions, hence more suppressed effects at low energy.

For the minimal EFT of Eq.~\eqref{Lag} studied in the letter, we identify $c_1=(M_{X_1}\Lambda_{X_1}^4)^{-1}+(M_{X_2}\Lambda_{X_2}^4)^{-1}$ at $\mu_1=M_{X_1}$, while all other $c_i,\bar c_i=0$. Eq.~\eqref{Lnnnum} then allows us to translate the $\Lnn$ values corresponding to the benchmark $\taunn$'s in Eqs.~\eqref{Lnnreach-current} and~\eqref{Lnnreach-future} into contours in the $(M_{X_1},\,\Lambda_{X_1})$ or $(M_{X_2},\,\Lambda_{X_2})$ plane, depending on which term gives the dominant contribution to $c_1$ (see Figs.~\ref{fig:LateDecay} and~\ref{fig:EarlyDecay} of the letter).

\subsection*{2.~Details of baryogenesis calculation}

\subsubsection{Boltzmann equations}

The Boltzmann equations to be solved for our minimal EFT are
\beq
\frac{dn_a}{dt} + 3 H n_a = C_a \qquad\qquad (a=1,2,3) \,,
\eeqn
where $n_{1,2}$ are the number densities of $X_{1,2}$, and $n_3$ represents $n_{B-L}$ ($n_B$) for $T>140$\,GeV ($T<140$\,GeV) when electroweak sphalerons are assumed to be active (inactive). We define
\beqa
\textstyle
W_{i_1 \dots i_m \to f_1 \dots f_n} &\equiv& 
\int (d\Pi_{i_1}\dots d\Pi_{i_m}) (d\Pi_{f_1}\dots d\Pi_{f_n}) \,(2\pi)^4\delta^4\bigl(\sum_\alpha p_{i_\alpha}-\sum_\beta p_{f_\beta}\bigr) \,\bigl(f_{i_1}^\text{eq}\dots f_{i_m}^\text{eq}\bigr) \,|\M_{i_1 \dots i_m \to f_1 \dots f_n}|^2 \CR
&=&
\int (d\Pi_{i_1}\dots d\Pi_{i_m}) (d\Pi_{f_1}\dots d\Pi_{f_n}) \,(2\pi)^4\delta^4\bigl(\sum_\alpha p_{i_\alpha}-\sum_\beta p_{f_\beta}\bigr) \,\bigl(f_{f_1}^\text{eq}\dots f_{f_n}^\text{eq}\bigr) \,|\M_{i_1 \dots i_m \to f_1 \dots f_n}|^2 
\,,\qquad
\eeqan
where $f^\text{eq}_a$ is the equilibrium distribution at zero chemical potential for species $a$. Assuming a common temperature is maintained for all species, we have
\beq
f_a = e^{\mu_a/T} f_a^\text{eq} \equiv r_a f_a^\text{eq} \equiv (1+\Delta_a) \,f_a^\text{eq} \,,
\eeqn
for the actual distribution of species $a$, with $\Delta_a$ characterizing the amount of departure from equilibrium. The collision terms can then be written in terms of the $W$'s and $r$'s,
\beqa
-C_1 &=& 
\bigl(r_u r_1 -r_d^2\bigr) \,W_{u X_1 \to \bar d \bar d}
+ \bigl(r_{\bar u} r_1 -r_{\bar d}^2\bigr) \,W_{\bar u X_1 \to d d}
+ \bigl(r_d r_1 -r_u r_d\bigr) \,W_{d X_1 \to \bar u \bar d}
+ \bigl(r_{\bar d} r_1 -r_{\bar u} r_{\bar d}\bigr) \,W_{\bar d X_1 \to u d}
\CR &&
+ \bigl(r_u r_1 -r_{\bar u} r_2\bigr) \,W_{u X_1 \to u X_2}
+ \bigl(r_{\bar u} r_1 -r_u r_2\bigr) \,W_{\bar u X_1 \to \bar u X_2}
+ \bigl(r_1 r_2 -r_u r_{\bar u}\bigr) \,W_{X_1 X_2 \to \bar u u}
\CR &&
+ \bigl(r_1 -r_{\bar u} r_{\bar d}^2\bigr) \,W_{X_1 \to u d d}
+ \bigl(r_1 -r_u r_d^2\bigr) \,W_{X_1 \to \bar u \bar d \bar d}
+ \bigl(r_1 r_u r_{\bar u} -r_2\bigr) \,W_{X_2 \to X_1 \bar u u}
\,,\\
-C_2 &=& 
\bigl(r_u r_2 -r_d^2\bigr) \,W_{u X_2 \to \bar d \bar d}
+ \bigl(r_{\bar u} r_2 -r_{\bar d}^2\bigr) \,W_{\bar u X_2 \to d d}
+ \bigl(r_d r_2 -r_u r_d\bigr) \,W_{d X_2 \to \bar u \bar d}
+ \bigl(r_{\bar d} r_2 -r_{\bar u} r_{\bar d}\bigr) \,W_{\bar d X_2 \to u d}
\CR &&
+ \bigl(r_{\bar u} r_2 -r_u r_1\bigr) \,W_{u X_1 \to u X_2}
+ \bigl(r_u r_2 -r_{\bar u} r_1\bigr) \,W_{\bar u X_1 \to \bar u X_2}
+ \bigl(r_1 r_2 -r_u r_{\bar u}\bigr) \,W_{X_1 X_2 \to \bar u u}
\CR &&
+ \bigl(r_2 -r_{\bar u} r_{\bar d}^2\bigr) \,W_{X_2 \to u d d}
+ \bigl(r_2 -r_u r_d^2\bigr) \,W_{X_2 \to \bar u \bar d \bar d}
+ \bigl(r_2 -r_1 r_u r_{\bar u}\bigr) \,W_{X_2 \to X_1 \bar u u}
\,,\\
-C_3 &=&
\bigl(r_u r_1 -r_d^2\bigr) \,W_{u X_1 \to \bar d \bar d}
- \bigl(r_{\bar u} r_1 -r_{\bar d}^2\bigr) \,W_{\bar u X_1 \to d d}
+ \bigl(r_u r_2 -r_d^2\bigr) \,W_{u X_2 \to \bar d \bar d}
- \bigl(r_{\bar u} r_2 -r_{\bar d}^2\bigr) \,W_{\bar u X_2 \to d d}
\CR &&
+ \bigl(r_d r_1 -r_u r_d\bigr) \,W_{d X_1 \to \bar u \bar d}
- \bigl(r_{\bar d} r_1 -r_{\bar u} r_{\bar d}\bigr) \,W_{\bar d X_1 \to u d}
+ \bigl(r_d r_2 -r_u r_d\bigr) \,W_{d X_2 \to \bar u \bar d}
- \bigl(r_{\bar d} r_2 -r_{\bar u} r_{\bar d}\bigr) \,W_{\bar d X_2 \to u d}
\CR &&
- \bigl(r_1 -r_{\bar u} r_{\bar d}^2\bigr) \,W_{X_1 \to u d d}
+ \bigl(r_1 -r_u r_d^2\bigr) \,W_{X_1 \to \bar u \bar d \bar d}
- \bigl(r_2 -r_{\bar u} r_{\bar d}^2\bigr) \,W_{X_2 \to u d d}
+ \bigl(r_2 -r_u r_d^2\bigr) \,W_{X_2 \to \bar u \bar d \bar d}
\CR &&
+ 2\,r_u r_d^2 \,W'_{u d d \to \bar u \bar d \bar d}
- 2\,r_{\bar u} r_{\bar d}^2 \,W'_{\bar u \bar d \bar d \to u d d}
\,,
\eeqan
where $W'_{u d d \to \bar u \bar d \bar d}$, $W'_{\bar u \bar d \bar d \to u d d}$ are computed from the corresponding matrix elements with contributions from on-shell $X_{1,2}$ exchange subtracted. We have grouped together terms that are identical as dictated by $CPT$ invariance, $W_{i\to f}=W_{\bar f\to\bar i}$ (where bar denotes $CP$ conjugate state).

To further simplify, we note that several processes conserve $CP$ up to one-loop level, and as a result
\beq
W_{d X_1 \to \bar u \bar d} = W_{\bar d X_1 \to u d} \,,\qquad
W_{d X_2 \to \bar u \bar d} = W_{\bar d X_2 \to u d} \,,\qquad
W_{X_1 \to u d d} = W_{X_1 \to \bar u \bar d \bar d} \,.
\eeqn
For the $CP$-violating processes, on the other hand, we define their $CP$-symmetric and $CP$-asymmetric components,
\beqa
&&
W_{u X_1 \to \bar d \bar d}^{(0)} \equiv \frac{1}{2}\, \bigl(W_{u X_1 \to \bar d \bar d} +W_{\bar u X_1 \to d d}\bigr) \,,
\qquad
\epsilon W_{u X_1 \to \bar d \bar d} \equiv \frac{1}{2}\, \bigl(W_{u X_1 \to \bar d \bar d} -W_{\bar u X_1 \to d d}\bigr) \,,
\CR &&
W_{u X_2 \to \bar d \bar d}^{(0)} \equiv \frac{1}{2}\, \bigl(W_{u X_2 \to \bar d \bar d} +W_{\bar u X_2 \to d d}\bigr) \,,
\qquad
\epsilon W_{u X_2 \to \bar d \bar d} \equiv \frac{1}{2}\, \bigl(W_{u X_2 \to \bar d \bar d} -W_{\bar u X_2 \to d d}\bigr) \,,
\CR &&
W_{u X_1 \to u X_2}^{(0)} \equiv \frac{1}{2}\, \bigl(W_{u X_1 \to u X_2} +W_{\bar u X_1 \to \bar u X_2}\bigr) \,,
\qquad
\epsilon W_{u X_1 \to u X_2} \equiv \frac{1}{2}\, \bigl(W_{u X_1 \to u X_2} -W_{\bar u X_1 \to \bar u X_2}\bigr) \,,
\CR &&
W_{X_2 \to u d d}^{(0)} \equiv \frac{1}{2}\, \bigl(W_{X_2 \to u d d} +W_{X_2 \to \bar u \bar d \bar d}\bigr) \,,
\qquad
\epsilon W_{X_2 \to u d d} \equiv \frac{1}{2}\, \bigl(W_{X_2 \to u d d} -W_{X_2 \to \bar u \bar d \bar d}\bigr) \,,
\CR &&
W_{u d d \to \bar u \bar d \bar d}^{\prime(0)} \equiv \frac{1}{2}\, \bigl(W'_{u d d \to \bar u \bar d \bar d} +W'_{\bar u \bar d \bar d \to u d d}\bigr) \,,
\qquad
\epsilon W_{u d d \to \bar u \bar d \bar d}^{\prime} \equiv \frac{1}{2}\, \bigl(W'_{u d d \to \bar u \bar d \bar d} -W'_{\bar u \bar d \bar d \to u d d}\bigr) \,.
\eeqan
As a consequence of $CPT$ invariance and unitarity, $\sum_f W_{i\to f} =\sum_f W_{\bar i\to f}$, which implies
\beq
-\epsilon W_{u X_1 \to \bar d \bar d} = \epsilon W_{u X_2 \to \bar d \bar d} = \epsilon W_{u X_1 \to u X_2} \,,
\qquad
\epsilon W_{X_2 \to u d d} = \epsilon W_{u d d \to \bar u \bar d \bar d}^{\prime} \,.
\eeq{unitarity}
Using these relations and noting that $r_d r_{\bar d} =1$ (because $\mu_d+\mu_{\bar d}=0$), the collision terms can be rewritten as
\beqa
-C_1 &=& 
\bigl(2\,W_{u X_1 \to \bar d \bar d}^{(0)} +2\,W_{d X_1 \to \bar u \bar d} +2\,W_{u X_1 \to u X_2}^{(0)} +\,W_{X_1 X_2 \to \bar u u} +2\,W_{X_1 \to u d d} +\,W_{X_2 \to X_1 \bar u u}\bigr)\,\Delta_1 
\CR &&
-\bigl(2\,W_{u X_1 \to u X_2}^{(0)} -\,W_{X_1 X_2 \to \bar u u} +\,W_{X_2 \to X_1 \bar u u} \bigr)\,\Delta_2
+\bigl(2\,\epsilon W_{u X_1 \to u X_2}\bigr) (\Delta_u+2\Delta_d)
\CR &&
+\bigl(W_{X_1 X_2 \to \bar u u}\bigr)\,\Delta_1\Delta_2
+\bigl(2\,\epsilon W_{u X_1 \to u X_2}\bigr)\, \Delta_2\Delta_u \,,
\\
-C_2 &=&
\bigl(2\,W_{u X_2 \to \bar d \bar d}^{(0)} +2\,W_{d X_2 \to \bar u \bar d} +2\,W_{u X_1 \to u X_2}^{(0)} +\,W_{X_1 X_2 \to \bar u u} +2\,W_{X_2 \to u d d}^{(0)} +\,W_{X_2 \to X_1 \bar u u}\bigr)\,\Delta_2 
\CR &&
-\bigl(2\,W_{u X_1 \to u X_2}^{(0)} -\,W_{X_1 X_2 \to \bar u u} +\,W_{X_2 \to X_1 \bar u u} \bigr)\,\Delta_1
-2\,\bigl(\epsilon W_{u X_1 \to u X_2} -\epsilon W_{X_2 \to u d d}\bigr) (\Delta_u+2\Delta_d)
\CR &&
+\bigl(W_{X_1 X_2 \to \bar u u}\bigr)\,\Delta_1\Delta_2
-\bigl(2\,\epsilon W_{u X_1 \to u X_2}\bigr)\, \Delta_1\Delta_u \,,
\\
-C_3 &=&
2\,\bigl(W_{u X_1 \to \bar d \bar d}^{(0)} +W_{u X_2 \to \bar d \bar d}^{(0)} +W_{d X_1 \to \bar u \bar d} +W_{d X_2 \to \bar u \bar d} +W_{X_1 \to u d d} +W_{X_2 \to u d d}^{(0)}\bigr) (\Delta_u+2\Delta_d)
\CR &&
+\bigl(2\,\epsilon W_{u X_1 \to u X_2}\bigr) (\Delta_2-\Delta_1)
-\bigl(2\,\epsilon W_{X_2 \to u d d}\bigr) \,\Delta_2
\CR &&
+\bigl(2\,W_{u X_1 \to \bar d \bar d}^{(0)}\bigr)\,\Delta_1\Delta_u
+\bigl(2\,W_{u X_2 \to \bar d \bar d}^{(0)}\bigr)\,\Delta_2\Delta_u
+\bigl(2\,W_{d X_1 \to \bar u \bar d}\bigr)\,\Delta_1\Delta_d
+\bigl(2\,W_{d X_2 \to \bar u \bar d}\bigr)\,\Delta_2\Delta_d \,.
\label{C3}
\eeqan
As $|\Delta_{u,d}|\ll1$ in all cases, we have only kept terms up to linear order in $\Delta_{u,d}$. In addition, we approximate $\Delta_{u,d}=e^{\mu_{u,d}/T}-1\simeq\mu_{u,d}/T$. We have dropped the $W_{u d d \to \bar u \bar d \bar d}^{\prime(0)}$ term, which is higher order in $1/\Lambda$.

Now the collision terms are written in terms of $\Delta_{1,2,u,d}$, while the LHS of the Boltzmann equations contain $n_{1,2,B(-L)}$. To relate the two sets of quantities, we note that, assuming Maxwell-Boltzmann distributions for $X_{1,2}$,
\beq
\Delta_a = \frac{n_a}{n_a^\text{eq}}-1 = n_a \,\biggl[\frac{M_a^2 T}{\pi^2}K_2(M_a/T)\biggr]^{-1} -1 \qquad\qquad (a=1,2)\,.
\eeqn
Meanwhile, the chemical potentials $\mu_{u,d}$ are related to $n_{B(-L)}$ (see e.g.~\cite{Buchmuller00}): for $T>140$\,GeV,
\beq
\Delta_u = -\frac{10}{79}\frac{n_{B-L}}{T^3} \,,\qquad
\Delta_d = \frac{38}{79}\frac{n_{B-L}}{T^3} \,,
\eeqn
as follows from equilibration of Yukawa interactions and $SU(3)$ and $SU(2)$ sphalerons, and conservation of hypercharge; for $T<140$\,GeV, 
\beq
\Delta_u = \biggl[2\,\frac{n_L}{T^3}+\bigl(2+N_d^{-1}+N_e^{-1}\bigr)\,\frac{n_B}{T^3}\biggr] \Bigl[1+\bigl(2+N_d^{-1}+3N_e^{-1}\bigr)\,N_u\Bigr]^{-1} \,,\qquad
\Delta_d = N_d^{-1}\biggl(3\,\frac{n_B}{T^3} -N_u\Delta_u\biggr)\,,
\eeqn
as follows from equilibration of charged current interactions, and conservation of electric charge and lepton number. Here $N_u$ ($N_d$, $N_e$) is the number of generations of relativistic up-type quarks (down-type quarks, charged leptons), and $n_L/T^3$ is a constant fixed by $-\frac{51}{79}\frac{n_{B-L}}{T^3}$ at $T=140$\,GeV.

Following the standard change of variables $x=M_{X_2}/T$, $Y_a=\frac{n_a}{s}=(\frac{2\pi^2}{45}h_\text{eff})^{-1}\frac{n_a}{T^3}$, we have
\beq
\frac{dY_a}{dx} = \biggl(\frac{\pi}{45}\biggr)^{1/2} \frac{M_\text{pl}M_{X_2}}{s^2\,x^2} g_*^{1/2}\, C_a \qquad\qquad (a=1,2,3) \,,
\eeqn
where $g_*^{1/2}=\frac{h_\text{eff}}{g_\text{eff}^{1/2}}\bigl(1+\frac{1}{3}\frac{T}{h_\text{eff}}\frac{dh_\text{eff}}{dT}\bigr)$. This is the final form of the Boltzmann equations that we use in our numerical solutions. In order to determine viable parameter space regions for baryogenesis, we set $\text{arg}(\eta_{X_1}^*\eta_{X_2}\eta_c)=\pi/2$ to maximize $CP$ violation, and look for solutions with the final $Y_B\ge 8.6\times10^{-11}$; for such parameter choices, the exact amount of observed baryon asymmetry can then be achieved with some suitable choice of $\text{arg}(\eta_{X_1}^*\eta_{X_2}\eta_c)\le\pi/2$.

As mentioned in the letter, if $X_2$ is sufficiently long-lived, its decay may dump significant entropy into the plasma, diluting the baryon asymmetry. We account for this effect by dividing the final $Y_B$ from solving the Boltzmann equations by a dilution factor $d_s=1.83\,h_\text{eff}^{1/4} M_{X_2} \, Y_{X_2}^\text{d}\,(\Gamma_{X_2} \, M_\text{pl})^{-1/2} = 1.42 \,x_\text{d}\, Y_{X_2}^\text{d}$, if $d_s>1$. Here $\Gamma_{X_2}$ is the total decay width of $X_2$, and $x_\text{d}$ and $Y_{X_2}^\text{d}$ are the values of $M_{X_2}/T$ and $Y_{X_2}$ at the time of $X_2$ decay, determined by $\Gamma_{X_2}=H$.

\subsubsection{Interaction rates}

We now provide analytical expressions for the interaction rates $W$ that appear in the collision terms. For a $2\to2$ process $ab\to cd$,
\beq
W_{ab\to cd} = n_a^\text{eq} n_b^\text{eq} \langle\sigma v\rangle_{ab\to cd} = \frac{T}{512\pi^5 S_i S_f} \int_{s_\text{min}}^\infty \frac{p_i p_f}{\sqrt{s}} \langle {\textstyle\sum} |\M|^2 \rangle_{ab\to cd} \,K_1(\sqrt{s}/T) \,ds \,,
\eeqn
where $S_i$, $S_f$ are symmetry factors for the initial and final states (e.g.\ $S_i=2$ if $a$ and $b$ are identical particles) and $p_i = |\overrightarrow{p_a}| = |\overrightarrow{p_b}|$, $p_f = |\overrightarrow{p_c}| = |\overrightarrow{p_d}|$ in the center of mass frame. The sum is over initial {\it and} final state spins and colors, while ``$\langle\,\rangle$'' means averaging over $\cos\theta$, with $\theta$ being the scattering angle in the center of mass frame. We take the upper limit of integration to $\infty$ for simplicity since the integrand is exponentially suppressed for center-of-mass energies above the EFT cutoff $\Lambda$ for temperatures where the EFT is valid ($T\ll\Lambda$). Computing the scattering amplitudes at tree level, we find
\beqa
p_i p_f \langle {\textstyle\sum} |\M|^2 \rangle_{u X_a \to \bar d \bar d} &=& \frac{1}{2}\, |\eta_a|^2 \bigl(s-M_a^2\bigr)^2 \bigl(s+2M_a^2\bigr) \,, \\
p_i p_f \langle {\textstyle\sum} |\M|^2 \rangle_{d X_a \to \bar u \bar d} &=& \frac{7}{2}\, |\eta_a|^2 \bigl(s-M_a^2\bigr)^2 \bigl(s+\textstyle\frac{1}{14}M_a^2\bigr) \,,\\
p_i p_f \langle {\textstyle\sum} |\M|^2 \rangle_{u X_1 \to u X_2} &=& |\eta_c|^2 \,s^{-3}\bigl(s-M_{X_1}^2\bigr)^2 \bigl(s-M_{X_2}^2\bigr)^2 \CR
&&\quad \Bigl[ s^2 +{\textstyle\frac{1}{8}}\bigl(M_{X_1}^2+M_{X_2}^2\bigr)s +{\textstyle\frac{1}{4}}M_{X_1}^2 M_{X_2}^2 +{\textstyle\frac{3}{4}}\cos2\phi_c M_{X_1} M_{X_2} \,s\Bigr] \,,\\
p_i p_f \langle {\textstyle\sum} |\M|^2 \rangle_{X_1 X_2 \to \bar u u} &=& \frac{1}{2}\, |\eta_c|^2 \Bigl[s^2 -2\bigl(M_{X_1}^2+M_{X_2}^2\bigr)s +\bigl(M_{X_2}^2-M_{X_1}^2\bigr)^2\Bigr]^{1/2}\cdot \CR
&&\quad \Bigl[s^2 -{\textstyle\frac{1}{2}}\bigl(M_{X_1}^2+M_{X_2}^2\bigr)s -{\textstyle\frac{1}{2}}\bigl(M_{X_2}^2-M_{X_1}^2\bigr)^2 -3\cos2\phi_c M_{X_1} M_{X_2} \,s\Bigr] \,,
\eeqan
where $\phi_c=\arg \eta_c$. We have seen above that all the $CP$ violation in $2\to2$ processes can be encoded in $\epsilon W_{u X_1 \to u X_2}$. Computing also one-loop diagrams for this process, we find
\beq
p_i p_f \langle {\textstyle\sum} |\M|^2 \rangle_{u X_1 \to u X_2} -p_i p_f \langle {\textstyle\sum} |\M|^2 \rangle_{\bar u X_1 \to \bar u X_2} = -\text{Im}\bigl(\eta_{X_1}^*\eta_{X_2} \eta_c\bigr) \frac{3}{32\pi s}M_{X_1} M_{X_2}\bigl(s-M_{X_1}^2\bigr)^2 \bigl(s-M_{X_2}^2\bigr)^2 \,.
\eeqn
We have explicitly checked that $CP$ violation in $u X_a \to \bar d \bar d$ satisfy expectations from unitarity relations Eq.~\eqref{unitarity}.

For decay processes,
\beq
W_{i\to abc} = n_i^\text{eq}\, \frac{K_1(M_{X_i}/T)}{K_2(M_{X_i}/T)}\, \Gamma_{X_i\to abc} = \frac{M_{X_i}^2 T}{\pi^2} K_1(M_{X_i}/T) 
\,\Gamma_{X_i\to abc} \,,
\eeqn
where $\Gamma$ is the rest frame decay rate, summed over final state spins and colors, and averaged over the initial state spin. 
At tree level, we have
\beqa
\Gamma_{X_a \to u d d} &=& \frac{3}{4096\pi^3} |\eta_a|^2 M_a^5\,, \\
\Gamma_{X_2 \to X_1 \bar u u} &=& \frac{1}{2048\pi^3} |\eta_c|^2 M_{X_2}^5 \Bigl\{\bigl(1-\rho^2\bigr)\Bigl[\bigl(1+\rho^2\bigr)\bigl(1-8\rho^2+\rho^4\bigr) +2\cos2\phi_c\,\rho\bigl(1+10\rho^2+\rho^4\bigr)\Bigr] \CR
&&\qquad\qquad\qquad\qquad -24\rho^3\Bigl[\rho -\cos2\phi_c\bigl(1+\rho^2\bigr)\Bigr]\log\rho\Bigr\} \,,
\eeqan
where $\rho=M_{X_1}/M_{X_2}$. The $CP$-violating decay rate at one-loop level reads
\beq
\Gamma_{X_2 \to u d d} -\Gamma_{X_2 \to \bar u \bar d \bar d} = \frac{1}{16384\pi^4} \text{Im}\bigl(\eta_{X_1}^*\eta_{X_2} \eta_c\bigr) M_{X_2}^7 \,\rho\Bigl[\bigl(1-\rho^4\bigr)\bigl(1-8\rho^2+\rho^4\bigr) -24\rho^4\log\rho\Bigr] \,.
\eeq{eGam2}
This function is maximized at $\rho=0.265\simeq 1/4$.

\subsubsection{Benchmark solutions}

To have a more detailed understanding of the baryogenesis scenarios discussed in the letter, let us examine a few benchmark solutions to the Boltzmann equations. We choose $M_{X_2}=4M_{X_1}=2\times10^5$\,GeV, which can accommodate solutions in both the late and the early decay scenarios, and consider the following three benchmarks:
\begin{itemize}
\item Degenerate: $\Lambda_{X_1}=\Lambda_{X_2}=\Lambda_c=1.5\times10^6$\,GeV.
\item Late decay: $\Lambda_{X_1}=1.5\times10^6$\,GeV, $\Lambda_{X_2}=\Lambda_c=80\,\Lambda_{X_1}$.
\item Early decay: $\Lambda_{X_2}=1.5\times10^6$\,GeV, $\Lambda_{X_1}=10\,\Lambda_{X_2}$, $\Lambda_c=0.2\,\Lambda_{X_2}$.
\end{itemize}
All three benchmarks induce \nnbar oscillation at a level that is consistent with current constraints and may be within reach of future searches. 

\begin{figure}[t]
\centering
\includegraphics[width=\linewidth]{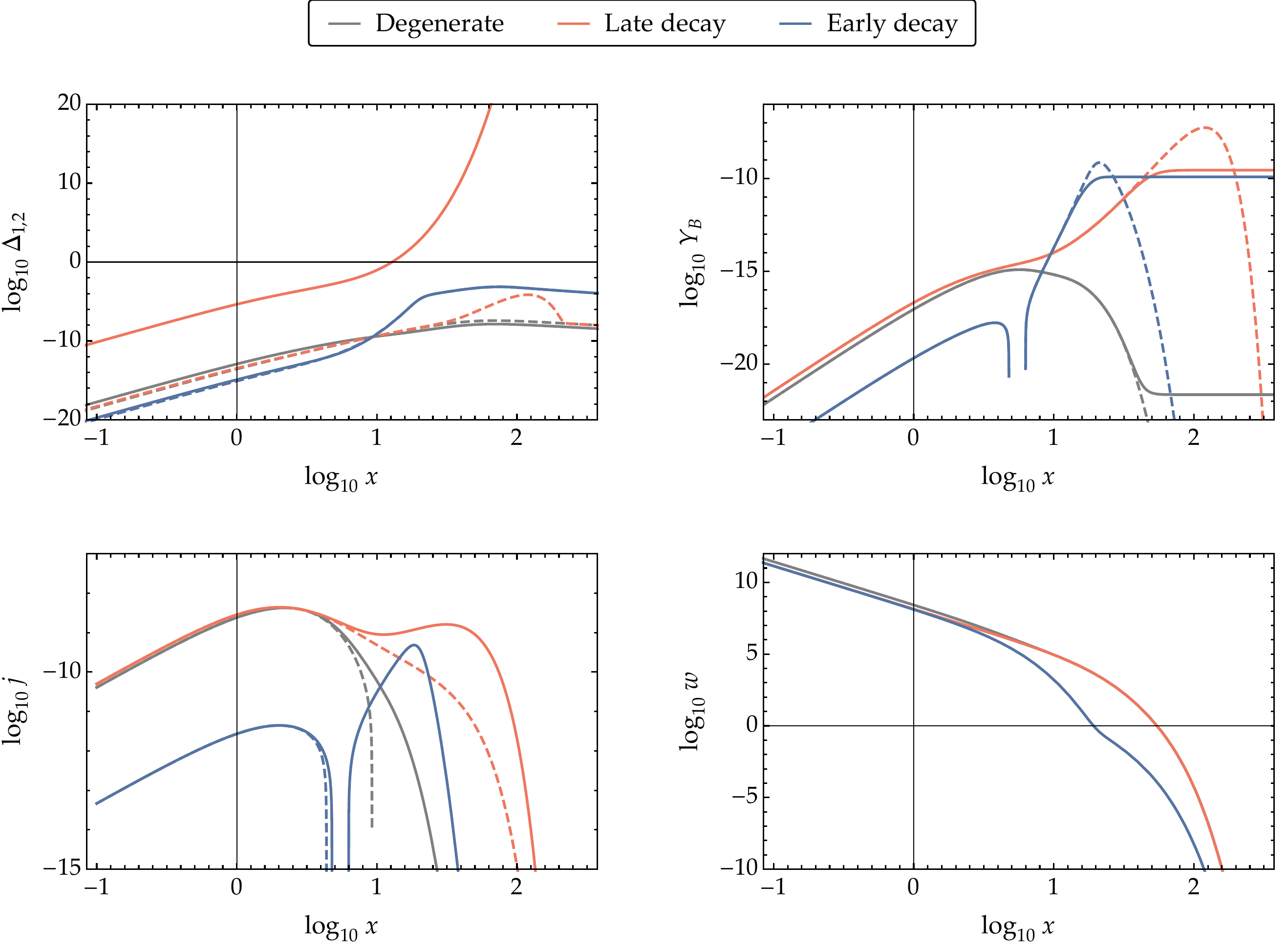}
\caption{\label{fig:sol}
Evolution of $\Delta_1=(Y_{X_1}-Y_{X_1}^\text{eq})/Y_{X_1}^\text{eq}$ (upper-left, dashed), $\Delta_2=(Y_{X_2}-Y_{X_2}^\text{eq})/Y_{X_2}^\text{eq}$ (upper-left, solid), and $Y_B$ (upper-right, solid) with $x=M_{X_2}/T$, in the three benchmark scenarios. $Y_B$ follows the adiabatic curve $Y_B^\text{adiabatic}(x)=j(x)/w(x)$ (upper-right, dashed) until $B$ violating interaction rates become Boltzmann suppressed. The source and washout functions $j(x)$ and $w(x)$ in the Boltzmann equation for $Y_B$, Eq.~\eqref{BEYB}, are plotted in the lower panels (solid). The source function $j(x)$ is dominated by contributions from $2\leftrightarrow2$ processes (lower-left, dashed) at early times, and by contributions from decays at the time of baryon asymmetry generation.
}
\end{figure}

We plot the evolution of various quantities from solving the Boltzmann equations in Fig.~\ref{fig:sol}. The upper-left panel shows the amount of departure from equilibrium for $X_1$ (dashed) and $X_2$ (solid), quantified by $\Delta_a=(Y_a-Y_a^\text{eq})/Y_a^\text{eq}$, while the solid curves in the upper-right panel show the baryon asymmetry $Y_B$.

We first note that, with the exception of $\Delta_2$  in the late decay benchmark, departures from equilibrium are always very small due to efficient depletion of $X_{1,2}$ number densities by rapid decays once they become nonrelativistic. As a rough estimate, assuming radiation domination, we have $\Gamma_{1\to3}/H\sim 10^{-5} \frac{M^5}{\Lambda^4}\frac{M_\text{pl}}{T^2}>10^{-5}\frac{M^3 M_\text{pl}}{\Lambda^4}$ when $T<M$, where $10^{-5}$ is the size of the phase space factor. For $M\sim 10^5$\,GeV and $\Lambda\sim 10^6$\,GeV, we have $\Gamma_{1\to3}/H \gtrsim 10^5$ and thus efficient decays that keep $\Delta_a\ll1$. On the other hand, $\Delta_2$  in the late decay benchmark evades this pattern with much higher values for $\Lambda_{2,u}\sim10^8$\,GeV, which result in $\Gamma_{1\to3}/H\sim10^{-3}(M_{X_2}/T)^2$, and thus later decay, for $X_2$. In this case, $\Delta_2$ starts to grow exponentially once the most efficient $2\to2$ process $d X_2\to\bar u\bar d$ freezes out, which happens when $\Gamma_{2\to2}/H\sim n_d^\text{eq} \langle\sigma v\rangle/H\sim g_\text{eff} \frac{T^3}{\pi^2}\cdot 10^{-2}\frac{M^2}{\Lambda^4}\cdot\frac{M_\text{pl}}{T^2}\sim 10 \,(T/M_{X_2})\sim1$, i.e.\ when $x=M_{X_2}/T\sim10$.

It is also worth noting that departures from equilibrium $\Delta_{1,2}$ are nonzero even when $X_{1,2}$ are relativistic, as a result of Hubble expansion. To see this, we write both sides of the Boltzmann equation for $X_a$ schematically as
\beqa
\text{LHS} &=& \frac{dY_a}{dx} = Y_a^\text{eq}\frac{d\Delta_a}{dx} + (1+\Delta_a)\frac{dY_a^\text{eq}}{dx} \simeq \frac{dY_a^\text{eq}}{dx} \sim -\frac{M}{T} \,,\\
\text{RHS} &\sim&\frac{M_\text{pl}}{MT^4}C_a = \frac{M_\text{pl}}{MT^4} \bigl(C_a^{2\to2}+C_a^{1\to3}\bigr) \sim -\frac{M_\text{pl}}{MT^4} \biggl( \frac{T^8}{\Lambda^4} +\frac{M^6T^2}{\Lambda^4} \biggr)\Delta_a \sim -\frac{M_\text{pl}T^4}{M\Lambda^4}\Delta_a \,.
\eeqan
Setting them equal, we have
\beq
\Delta_a \sim \frac{M^2\Lambda^4}{M_\text{pl}T^5} \sim x^5 \qquad (x\ll1)\,.
\eeqn
This power law dependence is clearly seen from the upper-left panel of Fig.~\ref{fig:sol}. Also note that $\Delta_a$ is larger for higher $\Lambda$, as it is harder to catch up with Hubble expansion when interactions are weaker. When nondegenerate $\Lambda$'s are involved in the $X_{1,2}$ number changing processes, the lowest of them tends to determine the total interaction rate, and thus $\Delta_a$. For example, at high temperatures, $\Delta_{1,2}$ are lower in the early decay benchmark compared to the degenerate case because of a lower $\Lambda_c$. They exceed the degenerate curves later when coannihilation becomes Boltzmann suppressed; from here on, $\Delta_1$ tends to grow faster due to a higher $\Lambda_{X_1}$, while the lower $\Lambda_c$ maintains $\Delta_1\simeq\Delta_2$ via $u\, X_1\leftrightarrow u\, X_2$, $X_2\leftrightarrow X_1\, \bar u\, u$ processes.

Next, to understand the trend of the $Y_B$ curves in the upper-right panel of Fig.~\ref{fig:sol}, it is useful to note that the Boltzmann equation for $Y_B$ has the following form,
\beq
\frac{dY_B}{d\log x} = j(x) -w(x)\, Y_B \,.
\eeq{BEYB}
Here $j(x)$ is a source function that is proportional to $CP$ violating interaction rates and departures from equilibrium,
\beq
j(x) \propto \bigl(2\,\epsilon W_{u X_1 \to u X_2}\bigr) (\Delta_2-\Delta_1)
-\bigl(2\,\epsilon W_{X_2 \to u d d}\bigr) \,\Delta_2 \,,
\eeq{jx}
see Eq.~\eqref{C3}. $w(x)$ is a washout function that tends to erase the baryon asymmetry. At high temperatures, both $j(x)$ and $w(x)$ vary slowly (as powers of $x$ as we will see below), so that the evolution of $Y_B$ is approximately adiabatic,
\beq
Y_B \simeq Y_B^\text{adiabatic} = j(x)/w(x) \qquad \text{(high $T$)}\,.
\eeqn
The adiabatic solutions are shown by the dashed curves in the upper-right panel of Fig.~\ref{fig:sol}. The true solutions follow the adiabatic approximation as long as $|\frac{d(j/w)}{d\log x}|\ll j$, which is seen to be the case until $x\sim10^{1\div2}$; after that, $j/w$ varies too fast for $Y_B$ to follow, and $Y_B$ freezes out.

We see from the plot that the key to generating sufficient baryon asymmetry in both late and early decay scenarios is the appearance of a sharp peak in $j(x)/w(x)$. This allows $Y_B$ to rise to significantly higher values compared to the degenerate case, and freeze out just before $j/w$ turns over. To see how the peak arises in each case, we plot the functions $j(x)$, $w(x)$ in the lower panels of Fig.~\ref{fig:sol} (solid curves). In addition, to compare contributions from $2\leftrightarrow2$ vs.\ $1\leftrightarrow3$ processes, we plot the former (corresponding to the first term on the RHS of Eq.~\eqref{jx}) with dashed curves in the lower-left panel. Quite generally, the ratio of the two scales as $T/M$ to some positive power, so $2\leftrightarrow2$ ($1\leftrightarrow3$) processes dominate at high (low) temperatures. This makes it clear that $1\leftrightarrow3$ processes are responsible for sufficient baryon asymmetry generation in both late and early decay scenarios.

To have a more detailed understanding of these plots, we first note that at high temperatures, 
\beqa
T\gg M: \qquad
j &\sim& \frac{M_\text{pl}}{T^5} \bigl(\epsilon W_{u X_1 \to u X_2}\bigr) (\Delta_2-\Delta_1)
\sim \frac{M_\text{pl}}{T^5} \frac{M^2 T^8}{\Lambda^6} \frac{M^2\Lambda^4}{M_\text{pl}T^5}
\sim \frac{M^4}{\Lambda^2 T^2} \sim x^2 \,,\\
w &\sim& \frac{M_\text{pl}}{T^5} W_{2\to2}
\sim \frac{M_\text{pl}}{T^5} \frac{T^8}{\Lambda^4} 
\sim \frac{M_\text{pl}T^3}{\Lambda^4} \sim x^{-3} \,,
\eeqan
and therefore,
\beq
Y_B \simeq j/w \sim \frac{M^4\Lambda^2}{M_\text{pl}T^5} \sim x^5 \qquad (x\ll1)\,.
\eeqn
These power law behaviors can be clearly seen in Fig.~\ref{fig:sol}. Note that in the parameter space probed by \nnbar oscillation, it is not possible to fully generate the observed baryon asymmetry while $X_{1,2}$ are relativistic (however, viable baryogenesis via $2\to2$ processes at $T>M$ is possible with higher $\Lambda$'s, as demonstrated in~\cite{Baldes1410}).

As the temperature falls below the $X_{1,2}$ masses, interaction rates become Boltzmann suppressed. From Eq.~\eqref{C3},
\beqa
T\ll M: \qquad\qquad
j &\sim& \frac{M_\text{pl}}{T^5} \cdot n_2^\text{eq}\frac{M^7}{\Lambda_{X_1}^2\Lambda_{X_2}^2\Lambda_c^2}\cdot \Delta_2 \,,\\
w &\sim& \frac{M_\text{pl}}{T^5} \biggl[ n_1^\text{eq}\frac{M^5}{\Lambda_{X_1}^4} \max(1,\Delta_1) + n_2^\text{eq}\frac{M^5}{\Lambda_{X_2}^4} \max(1,\Delta_2) \biggr] \,.
\eeqan
In the absence of exponential growth of $\Delta_{1,2}$, $j(x)\propto n_2^\text{eq}\propto e^{-M_{X_2}/T}$ simply falls exponentially when $T<M_{X_2}$. Meanwhile, when $\Lambda_{X_1}\sim\Lambda_{X_2}$, $w(x)$ is dominated by the term proportional to $n_1^\text{eq}\propto e^{-M_{X_1}/T}$, and so becomes exponentially suppressed at a later time when $T<M_{X_1}$. This results in a period of efficient washout of the baryon asymmetry generated previously, and ultimately, $j(x)/w(x)\propto e^{-(M_{X_2}-M_{X_1})/T}$. As we see from Fig.~\ref{fig:sol}, the degenerate benchmark curves indeed follow these expectations. 

In contrast, the late decay scenario features an exponentially growing $\Delta_2$, due to freeze-out of the $X_2$ abundance discussed above. This enhanced departure from equilibrium induces a plateau in the source function $j(x)$, before $X_2$ finally decays. Meanwhile, washout is still dominated by processes involving $X_1$, and $w(x)$ is similar to the degenerate case. The overall effect is thus a much higher $j(x)/w(x)$ than the degenerate case, peaked around $x\sim10^{2.1}$, allowing for sufficient baryon asymmetry generation.

The early decay scenario, on the other hand, features a dip in the washout function. With $\Lambda_{X_1}\gg\Lambda_{X_2}$, $w(x)$ is now dominated by the term proportional to $n_2^\text{eq}\propto e^{-M_{X_2}/T}$ at high temperatures, which falls off exponentially earlier than the $n_1^\text{eq}$ term. Baryon asymmetry generation is further assisted by the delayed (eventual) fall-off of the source function $j(x)$ due to the growth of $\Delta_2$ around $x\sim10$ explained previously. This also results in a gap between the ``$1\leftrightarrow3$ peak'' and the ``$2\leftrightarrow2$ peak'' of $j(x)$, as the latter falls off around the same time as in the degenerate benchmark because $\Delta_2-\Delta_1$ remains small. Overall, $j(x)/w(x)$ is significantly boosted compared to the degenerate case, with a peak around $x\sim 10^{1.3}$, corresponding to efficient baryogenesis at a time earlier than the late decay scenario.

\end{document}